\documentclass[preprintnumbers,preprint,aps,amsmath,amssymb]{revtex4}
\usepackage{epsfig}
\def\lsim{\:\raisebox{-1.1ex}{$\stackrel{\textstyle<}{\sim}$}\:}
\def\gsim{\:\raisebox{-1.1ex}{$\stackrel{\textstyle>}{\sim}$}\:}
\def\10{$SO(10)$}

\def\21{SU(2) $\otimes$ U(1) }

\def\422{$SU(4) \otimes SU(2) \otimes SU(2)$}
\def\321{SU(3) $\otimes$ SU(2) $\otimes$ U(1)}
\def\lsim{\raise0.3ex\hbox{$\;<$\kern-0.75em\raise-1.1ex\hbox{$\sim\;$}}}
\def\gsim{\raise0.3ex\hbox{$\;>$\kern-0.75em\raise-1.1ex\hbox{$\sim\;$}}}
\def\vev#1{\left\langle #1\right\rangle}

\newcommand{\ba}{\begin{array}}
\newcommand{\ea}{\end{array}}
\newcommand{\be}{\begin{equation}}
\newcommand{\ee}{\end{equation}}
\newcommand{\beqa}{\begin{eqnarray}}
\newcommand{\eeqa}{\end{eqnarray}}
\relax
\def\321{$SU(3)\times SU(2)\times U(1)$}

\begin{document}
\bigskip
\title[]{Fermion number conservation and two Higgs doublet models without tree level
flavour changing neutral currents}
\author{
Anjan S.  Joshipura\footnote{anjan@prl.res.in} and Bhavik P. 
Kodrani\footnote{bhavik@prl.res.in}}
\affiliation{
Physical Research Laboratory, Navarangpura, Ahmedabad 380 009, India 
\vskip 2.0truecm}
\begin{abstract}
\vskip 1.0 truecm
The charged fermion mass matrices are always invariant under $U(1)^3$ symmetry linked to the fermion number
transformation. A class of two Higgs doublet models (2HDM) can be identified by requiring that the definition of this symmetry 
in an arbitrary weak basis be independent of Higgs parameters such as the ratio of the Higgs vacuum expectation values. The tree level flavour changing neutral currents normally present in 2HDM are absent in this class of models but unlike the type I or type II Higgs doublet models, the charged Higgs couplings in these models contain additional flavour dependent CP violating phases. These phases can account for the recent hints of the beyond standard model CP violation in the   $B_d$ and $B_s$ mixing. In particular, there is a range of parameters in which new  phases do not contribute to the $K$ meson
CP violation but give identical new physics contribution to the  $B_d$ and $B_s$ meson mixing. Specific model realizations of the above scenario are
briefly discussed.  
\end{abstract}

\maketitle
\section{Introduction}
Observations at B-factory and Tevatron have established the Cabibbo Kobayashi Maskawa (CKM ) mechanism as the dominant source of CP violation. At the same time they have also thrown hints \cite{lenz,cdf,d0} of departures from the CKM picture. Two noteworthy hints are in CP violations in the $B_d$ and $B_s$
systems. Indirect determination of the CP asymmetry $S_{\psi K_S}$   using the fitted values of the CKM parameters \cite{sl} seems to differ from the direct measurement 
by about 1.7$\sigma$. Likewise, the CKM fits in the SM predict very small CP asymmetry $S_{\psi\phi }$ compared to large values based on observations by the CDF \cite{cdf} and 
the $D0$ \cite{d0} groups.  Global analysis by the UTfit group \cite{utfit} appear to show more than $3\sigma$  deviation from the SM prediction for 
the phase appearing in $S_{\psi\phi }$ . The corresponding result by the CKM fitter \cite{ckmfit} 
group is $2.5 \sigma$.  Both these features may point to the presence of  some additional CP violation. We wish to look here at the possibility of explaining such CP violation using the two Higgs doublet models (2HDM).

2HDM represent  the simplest  extension of SM involving only one additional doublet of Higgs. 
These models allow (1) flavour changing neutral currents (FCNC) at the tree level and (2) contain a charged Higgs which provides additional source of 
flavour violations. The most general 2HDM tend to generate large FCNC for moderate Higgs mass.
This has resulted in specific models (called type-I and type II 2HDM)  in which an additional discrete symmetry is  imposed to
eliminate the FCNC at the tree level. It turns out that the charged Higgs couplings do not provide any additional source of CP violation in these models. They cannot thus explain possibly large
CP violating phases in the neutral $B$ mixings if confirmed in future. One would need to go beyond the type-I and type-II 2HDM to explain new 
CP violating phases.
One  possibility perused in \cite{chengsher,asj,bgl,hw} is to invoke flavour symmetries to restrict structure of FCNC couplings rather than eliminating them altogether. This leads to models with suppressed FCNC and additional phases whose phenomenological consequences have been studied in 
\cite{bhavik1,bhavik2,bhavik3}. More economical possibility is to have models without tree level FCNC but additional phases in the charged Higgs couplings. Here we wish to discuss such models motivated by the studies of flavour symmetries of mass matrix in 2HDM. 

Flavour symmetries are often invoked in SM and beyond to restrict the structure of Yukawa couplings all of which cannot be directly determined from
experiments. Either one can impose some symmetry and study its consequences for fermion flavour structure or one can use experimental information to guess possible flavour symmetries
under which fermion mass matrices remain invariant.
Advantage of this approach is that  it directly relates the experimental observations to some symmetries of mass matrices. 
But relating symmetries of mass matrices to symmetry of Lagrangian is not straightforward in this bottom up approach.
Assumption  that symmetries of mass matrices form  sub-groups of the full symmetry at the Lagrangian level can lead to
identification of possible interesting flavour symmetries and this approach has been pursued in \cite{lam1,lam2}. 
General study of this approach, particularly, the relation between the structures of  mass matrices and symmetries enjoyed by them was recently made in \cite{lam2,gll}.  Lam  in his study \cite{lam1} of the neutrino mass matrix found that an arbitrary neutrino mass matrix $M_\nu$ can always be linked to a symmetry $S$ which leaves
$M_\nu$ invariant, $S^{T} M_\nu S=M_\nu$. It was then pointed out by Grimus, Lavoura and Ludl \cite{gll} that any Hermitian mass matrix $M_f M_f^\dagger$ obtained form a fermion mass matrix $M_f$ always possesses a symmetry $G_f=U(1)\times U(1)\times U(1)$
and the corresponding $G$ for the mass matrix of  the Majorana neutrinos  is $Z_2\times Z_2\times Z_2$. This is easy to prove. If $V$ is a unitary  matrix
which diagonalizes a Hermitian mass matrix $M_f M_f^\dagger$, i.e.
$$M_fM_f^\dagger =VD^2V^\dagger$$
with a diagonal $D$ then one can always construct an $S=VP(\alpha_i) V^\dagger$ which leaves $M_fM_f^\dagger $ invariant, i.e. $S^\dagger M_fM_f^\dagger S=M_fM_f^\dagger$.
$P(\alpha_i)$  refers to a diagonal phase matrix  with phases $\alpha_i$ and $S$ therefore generates a $U(1)\times U(1)\times U(1)$ symmetry.

The above reasoning can easily be  generalized to non Hermitian mass matrices. Define
\be  \label{slrdef}S_{L,R}=V_{L,R}  P(\alpha_i)V_{L,R}^\dagger~, \ee
where $V_{L,R}$ are unitary matrices diagonalizing a general non-Hermitian $M$
\be \label{dia} M=V_LDV_R^\dagger ~. \ee
It then follows that $S_{L,R}$ define a symmetry of $M$: 
\be \label{sym}  S_L^\dagger M S_R=M ~.\ee
The symmetries $S_{L,R}$ and the resulting form of $M$ may look complicated depending on the choice of $V_{L,R}$ but eq.(\ref{sym}) is  equivalent to the  statement of the  fermion number conservation of each generation by its mass term. This is trivial to see. 
Let $f_{L,R}$ denote the fermion fields in their  mass basis corresponding to a diagonal mass matrix $D$. In this basis, individual fermion number is conserved by the mass term:
\be \label{fermion} P^\dagger(\alpha_i)  D P(\alpha_i)=D~. \ee
Arbitrary weak basis would be defined as  $f'_{L,R}=V_{L,R}f_{L,R}$. The phase invariance of the mass term (eq.(\ref{fermion})) then manifests itself in the weak basis as invariance under $S_{L,R}$, eq.(\ref{sym})as can be seen by multiplying eq.(\ref{fermion}) by $V_L (V_R^\dagger )$ from left(right) and using eq.(\ref{dia}).  The Majorana mass terms for neutrinos do not
conserve fermion number but mass of each neutrino is $Z_2$ invariant which reflects as $Z_2\times Z_2\times Z_2$ symmetry discussed in \cite{gll}.

Eq.(\ref{sym})  remains true for any choice of  $V_{L,R}$.  This invariance thus holds for any choice of 
 mass matrix as emphasized in \cite{gll}. However, if one wishes to understand symmetries of mass matrices 
as arising from some flavour symmetries at the Lagrangian level then only specific class of 
symmetry transformations $S_{L,R}$ would  be admissible. It is desirable to specify $S_{L,R}$ a priori in this case and put some reasonable requirement on them. In this case mass matrix symmetries may have non-trivial content and can
restrict the 
structure of the allowed theories. This is made explicit below within 2HDM.
\section{Mass matrix symmetries and 2HDM}
 Two Higgs doublet models contain the following Yukawa couplings which provide sources of mass and additional flavour violations:
\be \label{yukawa1}
-{\cal L}_Y=\bar{Q}_{L}'(\Gamma_{1d}\phi_1+\Gamma_{2d}\phi_2)d_R'+\bar{Q}_{L}'(\Gamma_{1u}\tilde{\phi}_1+
\Gamma_{2u}\tilde{\phi}_2)u_R'+{\rm H.c.} ~,\ee
where $\Gamma_{iq}~~(i=1,2;q=u,d)$ are matrices in the generation space. $\phi_{1,2}$ denote  Higgs doublets and $\tilde{\phi}_i= i \tau_2\phi_i^*$. 
$Q'_{L}$ refer to three generations of doublet quarks and primed fields in the above  equation refer to various quark fields in the weak basis. The neutral component of a specific linear combination of the Higgs fields namely,
\be \label{phi}
\phi\equiv \cos\beta \phi_1+\sin\beta  e^{-i \theta}\phi_2\ee
is responsible for the mass generation
\be \label{mq} M_q=v (\cos\beta \Gamma_{1q}+\sin\beta \Gamma_{2q} e^{i\theta_q})=V_{qL} D_qV_{qR}^\dagger~,\ee
where 
$$\vev{\phi_1^0}=v\cos\beta~;~\vev{\phi_2^0}=v\sin\beta e^{i\theta}$$
$v\sim 174$ GeV  and $\theta_d=-\theta_u=\theta$.
The matrices $V_{qL,R}$ diagonalize $M_q$ and also determine its symmetry
\be \label{sqlr}S_{qL,R}=V_{q L,R}P_qV_{q L,R}^\dagger ~,\ee
\be \label{Sq}
S_{qL}^\dagger M_q S_{qR}=M_q~.\ee
$P_q$ is a diagonal phase matrix
$$P_q={\rm diag.}(e^{i\alpha_{1q}},e^{i\alpha_{2q}},e^{i\alpha_{3q}}) $$
In the most general situation, the matrices $S_{qL,R}$ generate two  independent  $U(1)\times U(1)\times U(1)$
symmetries $G_u$ and $G_d$ for the up and the down quarks mass matrices respectively. 
$G_u\times G_d$ invariance holds for arbitrary $M_u,M_d$ and specific $S_{qL,R}$ determined from them.
We put a mild requirement on
possible $S_{q L,R}$ namely that the form of $S_{q L,R}$ be independent of the 
parameters $\tan\beta$ and $\theta$ which are determined  entirely in the Higgs sector.
This innocuous requirement has important consequences.
Using the definition, eq.(\ref{mq}) of mass matrices and eq.(\ref{Sq}) , it immediately leads to
\be \label{gamaq}  S_{qL}^\dagger \Gamma_{iq}S_{qR}=\Gamma_{iq}~~~~~{i=1,2} ~. \ee
This shows that not  only the total mass matrix but the individual Yukawa couplings should also  respect the symmetry.
Let us parametrize $\Gamma_{iq}$ as
\be \label{paramet}
\Gamma_{iq}\equiv V_{qL} \tilde{\Gamma}_{iq}V_{qR}^\dagger~.\ee
Eqs. (\ref{sqlr},\ref{gamaq}) then imply
\be \label{sol}
P_q^\dagger\tilde{\Gamma}_{iq} P_q=\tilde{\Gamma}_{iq} ~. \ee
If $G_u,G_d$ refer to the full $U(1)\times U(1)\times U(1)$ symmetry with totally independent  $\alpha_{iq}$ then the  only non-trivial solution of
eq.(\ref{sol}) is a diagonal $\tilde{\Gamma}_{iq}$ for every $i$ and $q$. Yukawa couplings are then given as
\be \label{yukawa2}
\Gamma_{iq}=V_{qL} \gamma_{iq}V_{qR}^\dagger  ~,\ee
where $\gamma_{iq}$ are diagonal matrices with  complex entries. 
More general forms for $\tilde{\Gamma}_{iq}$ are allowed if one demands invariance with respect to subgroups of $G_u\times G_d$ and we will discuss this case in the next section.

Eq.(\ref{yukawa2}) has powerful phenomenological implications. To see these,  let us note that the Higgs combination orthogonal to one in eq.(\ref{phi}) namely,
\be \label{phif}
\phi_F\equiv -\sin\beta \phi_1+\cos\beta \phi_2 e^{-i\theta} \ee
generates all the Higgs induced flavour violations. The couplings of the neutral component $\phi_F^0$ are given as
\be \label{fcnc}
{-\cal L}_Y^0=\bar{q}_{L}F_q q_{R}\phi_F^0+{\rm H.c.} \ee
with
\begin{eqnarray} \label{fij}
F_q&\equiv& V_{qL}^\dagger (-\sin\beta \Gamma_{1q}+\cos\beta \Gamma_{2q} e^{i \theta_q})V_{qR}~,\nonumber \\
&=& (-\sin\beta \gamma_{1q}+\cos\beta \gamma_{2q} e^{i \theta_q})~,\end{eqnarray}
where we have used eq.(\ref{yukawa2}) to obtain the second line. It is seen that the FCNC matrix $F_q$ become diagonal along with the mass matrices and the tree level FCNC are absent. But the phases of $(F_q)_{ii}$ cannot be removed in the process of making the quark masses real and remain as physical parameters.
The charged component of $\phi_F$ corresponds to the physical charged Higgs field and its couplings are given in our case by
\be \label{hplus}
-{\cal L}_{H^+}=\frac{H^+}{v} \left( \bar{u}_{iL} V_{ij} (F_d)_{jj} d_{jR} - \bar{u}_{iR} V_{ij} (F_u^*)_{ii} d_{jL}\right)+{\rm H.c.} ~\ee

The above couplings  are similar to the charged Higgs couplings in 2HDM of type-I and II.
In those models, $(F_q)_{ii}$ are proportional to the corresponding quark masses $m_{iq}$ and are real.
Here $(F_q)_{ii}$ are general complex numbers which can provide  
new phases in the  $B_{d,s}-\bar{B}_{d,s}$ mixing.

\noindent Let us make several important remarks:\\
\noindent (1)  An interesting class of 2HDM without the tree level FCNC have been recently discussed in \cite{pich}. These models are obtained from general 2HDM by 
assuming that two Yukawa couplings $\Gamma_{1q}$ and $\Gamma_{2q}$ are proportional to each other. 
In the present case,  the two Yukawa coupling matrices are not proportional to each other but  the tree level FCNC are still absent.
The phases of $(F_q)_{ii}$ in the charged Higgs couplings are dependent on the flavour index $i$ unlike in models of \cite{pich}
which are characterized by a universal phases one for the up and the other for the down quarks.  If the diagonal matrices $\gamma_{1q}$ and $\gamma_{2q}$ in eq.(\ref{yukawa2}) are proportional then the present class of models reduce to the one in \cite{pich}.\\
\noindent (2) We are assuming in general that $S_{uL}\not = S_{dL}$. Such inequality can arise from some more fundamental flavour symmetry  once $SU(2)_L\times U(1)$ is broken. There are well-known specific examples. For example
$D_4$ symmetry \cite{d4} broken in a specific way leads to trivial phase symmetry for the (diagonal) charged lepton matrix and 
$\mu$-$\tau$ symmetry for the neutrino mass matrix. Similarly, $D_4$ \cite{lam2} and $A_4$  \cite{he} symmetries in the quark sectors also  are shown to lead to effectively different $S_{uL}$ and $S_{dL}$ for the quark mass matrices.\\
\noindent (3) If $S_{uL}\not = S_{dL}$ then neither the Yukawa interactions $(\ref{yukawa1})$ nor the charged current weak interactions remain invariant under 
symmetries of the mass matrices. This means that radiative corrections will not preserve \cite{lavoura} the structure implied by eq.(\ref{yukawa2}). This equation thus 
should  be regarded as a  means of identifying class of models without the tree level FCNC.
Just as in case of the 2HDM of type-I and type-II as well as the aligned models of \cite{pich}, the full Lagrangian of the present model is formally invariant under
the  fermion number transformation $q_{iL,R}\rightarrow e^{i\alpha_{qi}}q_{iL,R}$ accompanied by the change in the CKM matrix elements $V_{ij}\rightarrow e^{i \alpha_{iu}}V_{ij}e^{-i \alpha_{jd}}$. As a consequence of this all the radiative corrections in the model would display structure similar to the one obtained in the 
Minimal Flavour Violating \cite{mfv} models.\\
%
\section{Modeling the symmetries:}

We have used the  $G_u\times G_d$ symmetry of the quark mass matrices to identify models without the tree level FCNC. As already stated these symmetries
do not commute with the $SU(2)_L$ group and are   effective symmetries of the quark  mass matrices in general. We wish to discuss here two examples. In the first example, the symmetries $S_{uL}$ and $S_{dL}$ are identified  and thus can be imposed at the Lagrangian level. The other model provides a specific realization of the Yukawa alignment discussed in \cite{pich} and is a special case of the general 2HDM identified here.

Le us assume that 
$$S_{uL}=S_{dL}\equiv S_L$$
It then follows from the defining equation (\ref{sqlr}) that
\be \label{relation} P_u V=V P_d~.\ee
where $V=V_{uL}^\dagger V_{dL}$ defines the CKM matrix. $P_u, P_d$ are a priori independent phase matrices
generating $G_u\times G_d$. From the fact that the diagonal elements of $V$ are non-zero and O(1), one immediately concludes that $P_u=P_d$. Moreover, if all entries in $V$ are taken to be non-zero then one is further led to $P_u=P_d=I$ and the symmetry $S_{L}$  becomes trivial.
But since, the elements of $V$ are known to be hierarchical one may assume as a first approximation
\begin{equation}\label{vckm12}
V\approx \left( \begin{array}{ccc}
1&\lambda&0\\
-\lambda&1&0\\
0&0&1\end{array}\right)~,\ee
where $\lambda$ is the Wolfenstein parameter denoting approximately the Cabibbo angle.
This form is consistent with eq.(\ref{relation}) provided
\begin{eqnarray} \label{u1crossu1}
\alpha_{1d}&=&\alpha_{2d}=\alpha_{1u}=\alpha_{2u}\equiv \alpha~,\nonumber \\
\alpha_{3u}&=&\alpha_{3d}\equiv \eta ~. \end{eqnarray}
This relation defines a $U(1)\times U(1)$ symmetry with a non-trivial
$S_L$:
\be \label{SL}
(S_L)_{ij}=\delta_{ij}e^{i\alpha}+(e^{i \eta}-e^{i\alpha})( V_L)_{i3}(V_L)_{j3}^*~.\ee
Imposition of this symmetry would thus lead to approximately correct description of the quark mixing. Specifically, let us impose
$$ q_L'\rightarrow S_L q_L'~;~ q_{R}'\rightarrow S_{qR} q_{R}'~$$
as symmetries on the full Lagrangian. Here  $S_L$ is defined in eq.(\ref{SL}) and  $S_{qR}$ is obtained from it by the replacement $V_L\rightarrow V_{qR}$. 

The structure of the Yukawa couplings invariant under the above symmetry is given by: 
\begin{eqnarray}\label{gamma12}
\Gamma_{1u}=V_LV_{1u}\gamma_{1u} V_{Ru}^\dagger&;& \Gamma_{2u}=V_LV_{2u}\gamma_{2u} V_{Ru}^\dagger ~,\nonumber \\
\Gamma_{1d}=V_LV_{1d}\gamma_{1d} V_{Rd}^\dagger&;&\Gamma_{2d}=V_LV_{2d}\gamma_{2d} V_{Rd}^\dagger~, \end{eqnarray}
where $\gamma_{1q,2q}$ are diagonal matrices as before and $V_{1u,d}$ and $V_{2u,d}$ are independent matrices with a block diagonal structure having a non-trivial $12$ block. 
The mass matrices $M_q$ have simple structure in the basis defined by $q_L'\rightarrow \tilde{q}_L=V_L
q_L'$ and $q_R'\rightarrow \tilde{q}_R= V_{qR} q_R'$:
\be \label{12mq}
\tilde{M}_q=\left(\begin{array}{ccc}
X_q&A_q&0\\B_q&Y_q&0\\0&0&m_{3q}\end{array} \right) ~,\ee
where $m_{3u}=m_t~;~m_{3d}=m_b$. $\Gamma_{iq}$ also have similar structure in the same basis.
As expected, $\tilde{M}_q$ defined above leads to the CKM matrix as given in eq.(\ref{vckm12}). 

The above considerations do not dictate  any specific choice of $V_L$ and remain true for arbitrary $V_L$.  This may come from other independent 
considerations such as quark lepton unification. As an interesting example, let us assume that the $S_L$ defined as above also defines the symmetry of the neutrino
mass matrix in the flavour basis. Then $V_L$ can be related to the leptonic mixing matrix in which case one can choose 
to a good approximation
$(V_L)_{13}=0$ and $(V_L)_{23}=-(V_L)_{33}=-\frac{1}{\sqrt{2}}$.
Then
\be
S_L=\left(\begin{array}{ccc}
e^{i\alpha}&0&0\\
0&\frac{1}{2} (e^{i \eta}+e^{i\alpha})&-\frac{1}{2} (e^{i \eta}-e^{i\alpha})\\
0&-\frac{1}{2} (e^{i \eta}-e^{i\alpha})&\frac{1}{2} (e^{i \eta}+e^{i\alpha})\\ \end{array} \right) \ee
This corresponds to the generalized $\mu$-$\tau$ symmetry which exchanges the second and third generation fermions
 if $\alpha=0,\eta=\pi$.
This symmetry was earlier discussed in the context of quarks also\cite{dutta, asj2,bhavik1,gut}. In particular, it was shown in \cite{asj2} that one obtains the CKM matrix of the form given in eq.(\ref{vckm12}) by imposing the generalized $\mu$-$\tau$ symmetry. 
The discussion presented here shows that this result is not specific to the $\mu$-$\tau$ symmetry but would follow for any $U(1)\times U(1)$ symmetry as given in (\ref{u1crossu1}) with an arbitrary $V_L$. In this approach, $V_{ub},V_{cb}$ can arise from the small breaking of the $\mu$-$\tau$ symmetry as discussed in details in
\cite{asj2}.

Above model provides a good example of the "bottom approach" in which starting with symmetries of mass matrices we were led to 
a symmetry of Lagrangian which leads to the approximately correct CKM matrix at zeroth order. But the imposed $U(1)\times U(1)$ sub-group
lacks the power of the full $G_u\times G_d$ symmetry of the mass matrix. This follows from eq.(\ref{gamma12}). If $V_{1q}=V_{2q}$ in this equation then 
$\Gamma_{1q},\Gamma_{2q}$ and $M_q$ all get diagonalized by $V_{qL}=V_LV_{1q}$ with the result that there are no FCNC as can be verified by 
substituting this $V_{qL}$  in the first line of  eq.(\ref{fij}) and using eq.(\ref{gamma12}).
Thus unlike $G_u\times G_d$ symmetry, one needs additional
alignments condition $V_{1q}=V_{2q}$ in order to eliminate the FCNC.  We now discuss alternative model where such alignment is in-built.

The model is based on an additional  softly broken $SU(2)_H$ symmetry acting on the Higgs fields.
$(\phi_1,\phi_2)$ are taken as doublets under $SU(2)_H$. We also introduced two SM singlets $\chi_q\equiv (\chi_{1q},\chi_{2q})$, 
$q=u,d$ each transforming as doublet under $SU(2)_H$. Finally we impose a $Z_2$ symmetry under which $\chi_d$ and $d_R'$ change sign. This ensures that only
$\chi_d$ couples to d quarks and $\chi_u$ to the up quarks. Yukawa couplings are allowed as dimension five operators below some high scale $\Lambda$ as in the
Froggatt Nielsen approach \cite{fg}:
\be \label{yukawa3}
-{\cal L}_Y=\frac{1}{\Lambda}\left[ \bar{Q}_L'\Gamma_d(\chi_{1d}\phi_2-\chi_{2d}\phi_1)d_R'-\bar{Q}_L'\Gamma_u(\chi_{1u}\tilde{\phi}_1+\chi_{2u}\tilde{\phi}_2)u_R'\right]~.
\ee
One can use supersymmetry or appropriate $U(1)$ symmetry to forbid
couplings involving $\chi_{q}^\dagger$.
Vacuum expectation values of  $\chi_d,\chi_u$ at a scale $\lesssim \Lambda$  leads to 2HDM  Yukawa couplings as in eq.(\ref{yukawa1}) with the property
$$\Gamma_{2d}=-\frac{\vev{\chi_{1d}}}{\vev{\chi_{2d}}}\Gamma_{1d}=\frac{\vev{\chi_{1d}}}{\Lambda} \Gamma_d$$
$$\Gamma_{2u}=\frac{\vev{\chi_{2u}}}{\vev{\chi_{1u}}}\Gamma_{1u}=-\frac{\vev{\chi_{2u}}}{\Lambda} \Gamma_u$$
This relation realizes the alignments hypothesis in \cite{pich} and leads to models without the tree level FCNC. Since this is a subset of more general solution, eq.(\ref{yukawa2}) allowed by the $G_u\times G_d$ symmetry
the Yukawa couplings and the mass matrix $M_q$ remain invariant under this symmetry. The $SU(2)_H$ symmetry needs to be broken softly by mass terms in the Higgs sector to obtain the general vacuum structure.

Top quark Yukawa couplings in the above example also get suppressed by the the Froggatt Nielsen factor $\frac{\vev{\chi_{u}}}{\Lambda}$. This may not be desirable. This is avoided by adding
a third Higgs doublet $\phi_3$ instead of $\chi_{1u,2u}$. The $\phi_3$ is taken as singlet under $SU(2)_H$ and one imposes $\phi_3\rightarrow -\phi_3,u_{iR}'-\rightarrow u_{iR}'$. In this case the up quarks get their masses only from $\phi_3$ and the down quark Yukawa couplings remain the same as in the eq.(\ref{yukawa3}). One gets Yukawa alignment in the down quark sector as before.  Now the model has one more charged Higgs field which will mix with $H^+$ entering  Eq.(\ref{hplus}). If one denotes
now the lighter charged Higgs as $H^+$ then its couplings are obtained 
by the replacement $(F_d)_{ii}\rightarrow \eta_F (F_d)_{ii}~,(F_u)_{ii}\rightarrow m_{iu}\eta_3$ in eq.(\ref{hplus}). Here $m_{iu}$ denote the 
up quark masses and $\eta_F~(\eta_3)$ denotes the mixing of $H^+$ with $\phi_F^+(\phi_3^+)$. 
\section{Neutral Meson mixing}
As an example of the phenomenological application of the model, we discuss the neutral meson mixing induced by the charged 
Higgs couplings in eq.(\ref{hplus}). Some other application of this scheme are discussed in \cite{pich,similar}. We only discuss salient features here and defer a detailed discussion to a later publication \cite{bhavik4}. The 
$(F_d)_{ii}$ and $(F_u)_{ii}$ entering the $H^+$ couplings are determined by the diagonal Yukawa couplings $\gamma_{iq}$ which also determine 
corresponding quark masses, see eq.(\ref{yukawa2}). Let us make a simplifying assumption that the first two generation quark masses and the corresponding $(F_q)_{ii}$ are small compared to the third generation masses and  $(F_q)_{33}$. Retaining only the latter, eq.(\ref{hplus}) reduces to
\be \label{approxhplus}
-{\cal L}_{H^+}\approx\frac{H^+}{v} \left( \bar{u}_{iL} V_{i3} (F_d)_{33}b_R - \bar{t}_{R} V_{3j} (F_u^*)_{33}d_{jL}\right)+{\rm H.c.} ~\ee
The charged Higgs contribution to the $K^0-\bar{K}^0$ mixing arise only from the second term. The phase in 
$(F_u)_{33}$ can be absorbed in the definition of $H^+$. As a result  the above Lagrangian does not generate any new CP violating phases 
in the $K$ meson mixing as long as $(F_q)_{jj}$ are neglected for $j=1,2$. In this limit, the additional $H^+$ contribution to the $K^0-\bar{K}^0$ mixing has the same structure as in the MFV scenario \cite{mfv}. The same limit however can lead to non-trivial phases in the $B_q-\bar{B}_q$
mixing since the charged Higgs exchanges in this case involve both $(F_d)_{33}$ and $(F_u)_{33}$ and their phases cannot be simultaneously removed. More interestingly, the above interaction (\ref{approxhplus}) distinguishes between the $d$ and $s$ quarks only through the CKM factor and not through additional phases. This results in strong correlations among the CP violation in $B_s$ and $B_d$ system.

Let us make the above remarks  more quantitative. The $B_q^0-\bar{B}_q^0$  ($q=d,s$) mixing amplitude can be parametrized in the presence of new physics contribution as follows:
\be \label{np}
\vev{B_q|{\cal H}^{SM}+{\cal H}^{NP}|\bar{B}_q}\equiv \vev{B_q|{\cal H}^{SM}|\bar{B}_q}(1+\kappa_q e^{\phi^{NP}_q})\equiv 
|\vev{B_q|{\cal H}^{SM}|\bar{B}_q}|\rho_q e^{-2 i (\beta_q+\phi_q)}~, \ee
where $\beta_q$ represent the relevant phase in case of the SM and $\phi_q$ are the charged Higgs induced phases.
The box diagrams involving $WH$ and $HH$ lead to new physics contribution involving the charged Higgs $H$.
The interaction in eq.(\ref{approxhplus}) lead to the following effective Hamiltonian at the weak scale \cite{bhavik4}:
\be \label{effective H}
{\cal H}^{NP}=C_1\bar{q_L}\gamma^\mu b_L \bar{q_L}\gamma_\mu b_L+C_2\bar{q_R} b_L \bar{q_R}b_L ~,\ee
where $C_{1,2}$ are the Wilson coefficients. Taking matrix element of the above equation and comparing with the 
SM result leads to \cite{bhavik4}
\be \label{kapaq}
\kappa_q e^{i\sigma_q}=\frac{4 \pi^2}{G_F^2 M_W^2\eta_B S_0(x_t)}\left[C_1-5/24 C_2 \left(  \frac{M_{B_q}}{m_b+m_q}\right)^2\frac{B_{2q}}{B_{1q}}\right]  ~,\ee
where $G_F,M_W$ respectively denote the Fermi coupling constant and the W boson mass. $S_0(x_t)\approx 2.3 $ for $m_t\sim 161 $ GeV.
$\eta_B\approx 0.55$ refers to the QCD correction to the Wilson operator in the SM,$B_{1q,2q}$ are the bag factors which enter the operator matrix elements in 
eq.(\ref{effective H}) and $M_{B_q},m_b,m_q$ respectively denote the masses for the $B_q$ mesons,b quark and the $d,s$ quarks for $q=d,s$.

The Wilson coefficients $C_{1,2}$ are independent of the flavour $q=d,s$ of the light quark in $B_q$. 
Mild dependence of $\kappa_q$ on $q$
arise from the operator matrix element multiplying $C_2$ in eq.(\ref{kapaq}).
This leads to two predictions: To a good approximation, ($i$) $\kappa_d \approx \kappa_s$
and ($ii$) $\phi_{d}^{NP}\approx \phi_s^{NP}$. This implies from eq.(\ref{np}) that
\be \label{ratio}
\frac{\Delta M_d}{\Delta M_s}\approx \frac{\Delta M_d^{SM}}{\Delta M_s^{SM}}~,\ee
where $\Delta M_q$ denote the values of the $B_q^0-\bar{B}_q^0$ mass difference in the presence of new physics.

Equality of $\kappa_d$ and $\kappa_s$ as well as $\phi_d^{NP}$ and $\phi_s^{NP}$ also imply through eq.(\ref{np}) 

\be \label{phases}
\phi_d\approx \phi_s~ \ee
The detailed phenomenological consequences of this prediction are already discussed in \cite{bg} in a model independent manner. It appears to be in the right direction for explaining the CP violating anomalies. In case of 
$B_d$, the unitarity triangle angle $\beta$
$$ \sin 2\beta=0.84\pm 0.09$$ as determined \cite{sl} using the information from $V_{cb},\epsilon_K$  and $\frac{\Delta M_d}{\Delta M_s}$ is found to be higher than
the value 
$$0.681\pm 0.025$$
obtained from the mixing induced asymmetry in $B\rightarrow J/\psi K_S$ decay. Since the latter measures $\beta+\phi_d$, the above information can be reconciled \cite{bg}  with a negative $\phi_d\approx -10^\circ$. $\phi_d\approx \phi_s$ then implies a sizable
asymmetry \cite{bg}  $S_{\psi\phi}=\sin 2(\beta_s-\phi_s)\sim 0.4 $ in $B_s$ decay as is indicated by the analysis of UTfit \cite{utfit} or the CKMfitter  \cite{ckmfit} group.
The charged Higgs induced CP violation as discussed here thus provides a concrete realization of the scenario proposed in
\cite{bg} and comes with verifiable predictions, eq.(\ref{ratio}) and eq.(\ref{phases}). The same type of scenario also follows if the CP violation comes through neutral scalar-pseudoscalar mixing \cite{bhavik1}.
At the quantitative level one can use the experimental observations to limit the allowed parameter space of the 2HDM of the type proposed here. This will be separately discussed \cite{bhavik4}.
\section{Note added:} While preparing the manuscript, we became aware of \cite{similar} which contains an ansatz for Yukawa couplings more general than the one reported in \cite{pich}.
 

\begin{thebibliography}{99}
\bibitem{lenz} A.~Lenz and U.~Nierste,
  arXiv:hep-ph/0612167; A.~J.~Lenz,
  arXiv:0808.1944 [hep-ph]; 
  Nucl.\ Phys.\ Proc.\ Suppl.\  {\bf 177-178}, 81 (2008)
  [arXiv:0705.3802 [hep-ph]];  U.~Nierste,
  arXiv:hep-ph/0612310; F.~J.~Botella, G.~C.~Branco and M.~Nebot,
Nucl.\ Phys.\  B {\bf 768}, 1 (2007)[arXiv:hep-ph/0608100];The Heavy Flavour Averaging Group (HFAG), http://www.slac.stanford.edu/xorg/hfag/osc/PDG2009/\# BETAS.
M.~Bona {\it et al.},
  arXiv:0906.0953 [hep-ph].
%
\bibitem{cdf}  T.~Aaltonen {\it et al.}  [CDF Collaboration],
  Phys.\ Rev.\ Lett.\  {\bf 100}, 161802 (2008)
  [arXiv:0712.2397 [hep-ex]].
%
\bibitem{d0} V.~M.~Abazov {\it et al.}  [D0 Collaboration],
  Phys.\ Rev.\ Lett.\  {\bf 101}, 241801 (2008)
  [arXiv:0802.2255 [hep-ex]].
%
\bibitem{sl}  E.~Lunghi and A.~Soni,
  arXiv:0903.5059 [hep-ph].
%
\bibitem{utfit} M.~Bona {\it et al.}  [UTfit Collaboration],
  arXiv:0803.0659 [hep-ph] and  http://www.utfit.org, UTfit Collaboration (M. Bona 
{\it et al}.)
%
\bibitem{ckmfit} O.~Deschamps,
  arXiv:0810.3139 [hep-ph] and   http://ckmfitter.in2p3.fr,  (J. Charles {\it et al}).
%
\bibitem{chengsher} T.~P.~Cheng and M.~Sher,
  Phys.\ Rev.\  D {\bf 35}, 3484 (1987).
%
\bibitem{asj} A.~S.~Joshipura,
  Mod.\ Phys.\ Lett.\  A {\bf 6}, 1693 (1991).
%
\bibitem{bgl} G.~C.~Branco, W.~Grimus and L.~Lavoura,
  Phys.\ Lett.\  B {\bf 380}, 119 (1996) [arXiv:hep-ph/9601383].
%
\bibitem{hw} A.~Antaramian, L.~J.~Hall and A.~Rasin,
  Phys.\ Rev.\ Lett.\  {\bf 69}, 1871 (1992)
  [arXiv:hep-ph/9206205]; L.~J.~Hall and S.~Weinberg,
  Phys.\ Rev.\  D {\bf 48}, 979 (1993)[arXiv:hep-ph/9303241].
%
\bibitem{bhavik1}  A.~S.~Joshipura and B.~P.~Kodrani,
  Phys.\ Lett.\  B {\bf 670}, 369 (2009)
  [arXiv:0706.0953 [hep-ph]].
  %
\bibitem{bhavik2} A.~S.~Joshipura and B.~P.~Kodrani,
  Phys.\ Rev.\  D {\bf 77}, 096003 (2008)
  [arXiv:0710.3020 [hep-ph]].
%
\bibitem{bhavik3} A.~S.~Joshipura and B.~P.~Kodrani,
  Phys.\ Rev.\  D {\bf 81}, 035013 (2010)
  [arXiv:0909.0863 [hep-ph]].
%
%
\bibitem{lam1} C.~S.~Lam,
  Phys.\ Rev.\  D {\bf 74}, 113004 (2006);
  [arXiv:hep-ph/0611017].
%
\bibitem{lam2} C.~S.~Lam,
  Phys.\ Rev.\ Lett.\  {\bf 101}, 121602 (2008)
  [arXiv:0804.2622 [hep-ph]];
  Phys.\ Rev.\  D {\bf 78}, 073015 (2008)
  [arXiv:0809.1185 [hep-ph]].
%
\bibitem{gll}  W.~Grimus, L.~Lavoura and P.~O.~Ludl,
  J.\ Phys.\ G {\bf 36}, 115007 (2009)
  [arXiv:0906.2689 [hep-ph]].
%
\bibitem{pich} A.~Pich and P.~Tuzon,
  Phys.\ Rev.\  D {\bf 80}, 091702 (2009)
  [arXiv:0908.1554 [hep-ph]].
%
\bibitem{d4} W.~Grimus and L.~Lavoura,
  Phys.\ Lett.\  B {\bf 572}, 189 (2003)
  [arXiv:hep-ph/0305046].
%
\bibitem{he}  X.~G.~He, Y.~Y.~Keum and R.~R.~Volkas,
  JHEP {\bf 0604}, 039 (2006)
  [arXiv:hep-ph/0601001].
%
\bibitem{lavoura} P.~M.~Ferreira, L.~Lavoura and J.~P.~Silva,
  arXiv:1001.2561 [hep-ph].
%
\bibitem{mfv}  G.~D'Ambrosio, G.~F.~Giudice, G.~Isidori and A.~Strumia,
  Nucl.\ Phys.\  B {\bf 645}, 155 (2002)
  [arXiv:hep-ph/0207036].
%
\bibitem{dutta} Y.~Koide, H.~Nishiura, K.~Matsuda, T.~Kikuchi and T.~Fukuyama,
  Phys.\ Rev.\  D {\bf 66}, 093006 (2002)
  [arXiv:hep-ph/0209333];
 K.~Matsuda and H.~Nishiura,
  Phys.\ Rev.\  D {\bf 69}, 053005 (2004)
  [arXiv:hep-ph/0309272];
  Phys.\ Rev.\  D {\bf 69}, 117302 (2004);
  [arXiv:hep-ph/0403008]; 
  Phys.\ Rev.\  D {\bf 71}, 073001 (2005);
  Phys.\ Rev.\  D {\bf 72}, 033011 (2005)
  [arXiv:hep-ph/0506192];
  Phys.\ Rev.\  D {\bf 73}, 013008 (2006)
  [arXiv:hep-ph/0511338];
 H.~Nishiura, K.~Matsuda and T.~Fukuyama,
  Int.\ J.\ Mod.\ Phys.\  A {\bf 23}, 4557 (2008)
  [arXiv:0804.4515 [hep-ph]];
 A.~Datta and P.~J.~O'Donnell,
  Phys.\ Rev.\  D {\bf 72}, 113002 (2005)
  [arXiv:hep-ph/0508314].
%
\bibitem{asj2} A.~S.~Joshipura,
  Eur.\ Phys.\ J.\  C {\bf 53}, 77 (2008)
  [arXiv:hep-ph/0512252].
%
\bibitem{gut} A.~S.~Joshipura, B.~P.~Kodrani and K.~M.~Patel,
  Phys.\ Rev.\  D {\bf 79}, 115017 (2009)
  [arXiv:0903.2161 [hep-ph]].
%
\bibitem{fg} C.~D.~Froggatt and H.~B.~Nielsen,
  Nucl.\ Phys.\  B {\bf 147}, 277 (1979).
%
\bibitem{similar}  Y.~H.~Ahn and C.~H.~Chen,
  arXiv:1002.4216 [hep-ph]
%
\bibitem{bhavik4} A.~S.~Joshipura and B.~P.~Kodrani, in preparation
%
\bibitem{bg} A.~J.~Buras and D.~Guadagnoli,
  Phys.\ Rev.\  D {\bf 78}, 033005 (2008)
  [arXiv:0805.3887 [hep-ph]].


\end{thebibliography}
\end{document}